\documentclass[aps,prc,twocolumn, superscriptaddress,showkeys,nofootinbib]{revtex4-1}  

\usepackage[utf8]{inputenc}

\usepackage{graphicx,color}
\usepackage{amsmath,amssymb,amsfonts}
\usepackage{hyperref}

\usepackage{xspace}	

\begin{document}
\title{
Understanding baryon stopping at the BNL Relativistic Heavy Ion Collider top energies
}
\medskip
\author{Niseem~Magdy} 
\email{niseem.abdelrahman@tsu.edu}
\affiliation{Department of Physics, Texas Southern University, Houston, TX 77004, USA}
\affiliation{Physics Department, Brookhaven National Laboratory, Upton, New York 11973, USA}
\author{Prithwish Tribedy}
\affiliation{Physics Department, Brookhaven National Laboratory, Upton, New York 11973, USA}
\author{Chun Yuen Tsang}
\affiliation{Department of Physics, Kent State University, Kent, Ohio, 44242, USA}
\affiliation{Physics Department, Brookhaven National Laboratory, Upton, New York 11973, USA}
\author{Zhangbu Xu}
\affiliation{Department of Physics, Kent State University, Kent, Ohio, 44242, USA}
\affiliation{Physics Department, Brookhaven National Laboratory, Upton, New York 11973, USA}
\begin{abstract}
The nucleon exhibits a rich internal structure governed by Quantum Chromodynamics (QCD), where its electric charge arises from valence quarks, while its spin and mass emerge from complex interactions among valence quarks, sea (anti-)quarks, and gluons. At the advent of QCD, an alternative hypothesis emerged suggesting, at high energies, the transport of a nucleon's baryon number could be traced by a non-perturbative configuration of gluon fields connecting its three valence quarks, forming a $Y$-shaped topology known as the gluon junction. Recent measurements by the STAR experiment are compatible with this scenario. 
In light of these measurements, this study aims to explore the mechanisms of baryon transport in high-energy nuclear collisions using the PYTHIA-8 framework, which incorporates a state-of-the-art hadronization model with advanced Color Flow (CF) and Color Reconnection (CR) mechanisms that mimic signatures of a baryon junction. 
Within this model setup, we investigate (i) the rapidity slope of the net-baryon distributions in photon-included processes ($\gamma$+p) and (ii) baryon over charge transport in the isobaric (Ru+Ru and Zr+Zr) collisions.
Our study highlights the importance of the CF and CR mechanisms in PYTHIA-8, which play a crucial role in baryon transport. The results show that the CF and CR schemes significantly affect the isobaric baryon-to-charge ratio, leading to different predictions for baryon stopping and underscoring the need to account for CF and CR effects in comparisons with experimental measurements.
\end{abstract}
\maketitle

\section{INTRODUCTION} \label{sec:2}
\begin{figure}[!h]
 \centering{
 \includegraphics[width=0.99\linewidth,angle=0]{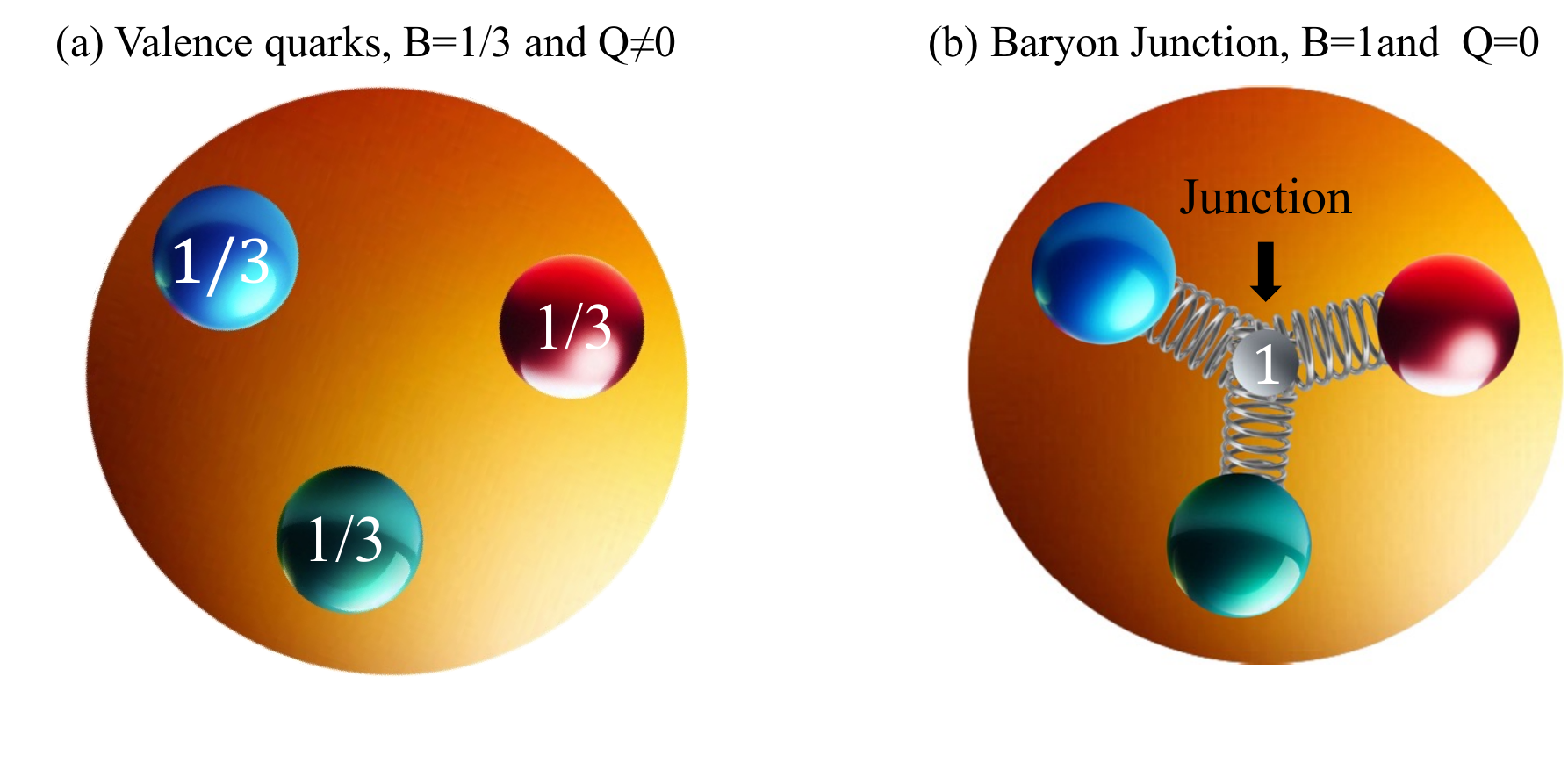}
\vskip -0.26cm
 \caption{
Schematic illustration of two types of baryon number carriers: panel (a) valence quarks and panel (b) baryon junction.
 \label{fig:00}
 }
 \vskip -0.26cm
 }
 \end{figure}

One of nature's fundamental principles is the conservation of baryon number, which plays a critical role in ensuring the stability of the proton, the lightest baryon known in nature. In the Standard Model, quarks are the fundamental constituents of hadrons, which carry electric charge \(Q\ne 0\). Each quark is assigned a baryon number of either $B = \pm \frac{1}{3}$, resulting in a total baryon number of one for baryons and zero for mesons (see Fig.\ref{fig:00} (a)).

In the early 1970s, an alternative picture was proposed in which, at high energies, the transport of baryon number \(B=1\) is traced by a \(Y\)-shaped structure connecting the three valence quarks (see Fig.~\ref{fig:00}(b)). This junction picture is theoretically appealing and preserves the gauge invariance of the baryon wave function~\cite{Artru:1974zn, Rossi:1977cy, Kharzeev:1996sq}.
Indeed, Lattice QCD calculations have explored the possibility of such a \(Y\)-shaped flux-tube structure in baryons~\cite{Suganuma:2004zx, Takahashi:2000te, Bissey:2006bz}, while string-inspired approaches have also been investigated in recent studies of the proton’s geometry in nuclear collisions~\cite{Coleman-Smith:2013rla} and in diffractive \(J/\psi\) production at HERA~\cite{Mantysaari:2016jaz, Xiang:2024zyu}. 
Experimental confirmation of a \(Y\)-shaped configuration inside a proton, and whether it is responsible for transporting baryon number instead of the valence quarks, is challenging. This difficulty arises because the observable signatures of quark-based and junction-based configurations are nearly indistinguishable in most physical processes.

 \begin{figure*}[t]
 \centering{
 \includegraphics[width=0.9 \linewidth,angle=0]{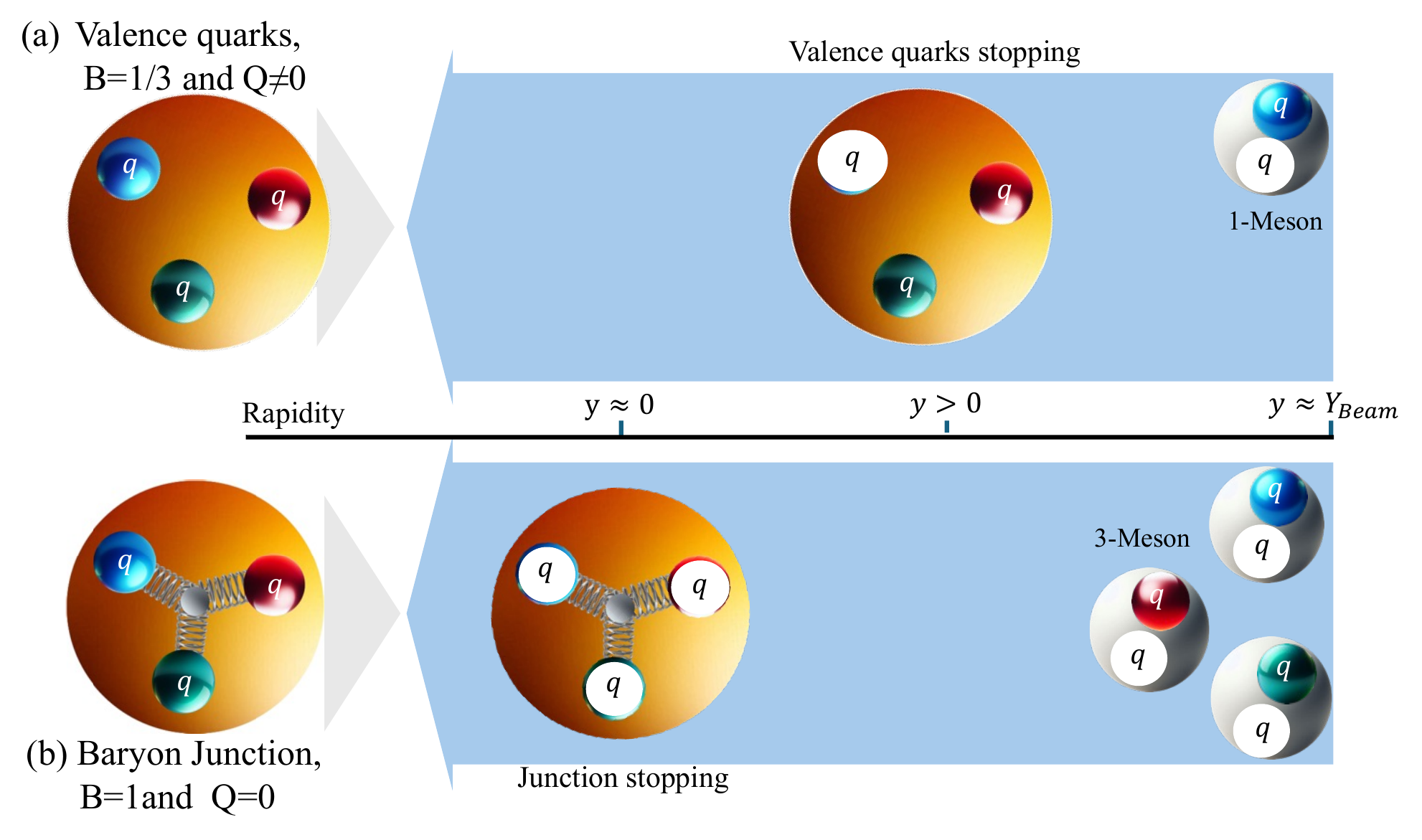}
\vskip -0.26cm
 \caption{
 Schematic illustration of the baryon number carriers and their transport mechanisms based on~\cite{STAR:2024lvy}. On the left: a depiction of the case when valence quarks carry the baryon number. On the right is a representation of the case when the baryon junction carried the baryon number. Unstopped valence quarks (solid spheres) can form mesons near beam rapidity. Open spheres represent produced quarks and anti-quarks.
 \label{fig:00x}
 }
 \vskip -0.16cm
 }
 \end{figure*}

Recent developments have proposed ways to disentangle these two pictures by carefully analyzing the interplay between quark stopping and junction stopping, which can yield distinct signatures in baryon production and composition~\cite{Lewis:2022arg, Frenklakh:2023pwy, Lv:2023fxk, Frenklakh:2024mgu, STAR:2024lvy, Magdy:2024dpm, Tsang:2024zsq}. The baryon-junction mechanism produces baryons \(B\) with quark compositions different from the colliding baryons ($p$ or $n$) and associated pions, contributing to high-multiplicity events at forward rapidity~\cite{Kharzeev:1996sq}. This feature helps distinguish valence-quark from gluon-junction transport, as shown in Fig.~\ref{fig:00x}. Additionally, the slow exponential dependence of baryon cross-section with rapidity loss ($y-Y_{\rm beam}$), characteristic of junction-stopping processes, favors the baryon junction model over the valence quark model in high-energy collisions.
For example, in a semi-inclusive photon-induced process involving a proton, such as $\gamma^* + p \rightarrow B + X$, the rapidity distribution of the produced net baryon is expected to follow~\cite{Kharzeev:1996sq}:
\begin{equation}\label{Eq:1}
\left. \dfrac{dN}{dy} \right|_{Net-B} \approx \exp(-\alpha_J(y-Y_{\rm beam})),
\end{equation}
where the exponent \(\alpha_J\), characterizing the rapidity slope, is determined by the baryon junction’s Regge intercept \(\alpha_n^J\).
In this picture, $\alpha_n^J=-0.5,0,0.5$ is determined by how many valence quarks ($n=$2, 1, or 0) from the colliding proton $p$ accompany the junction during baryon ($B$) formation, along with the creation of ($3-n$) associated mesons ($X$). This results in a small effective rapidity slope, which can serve as signatures of the baryon junction-stopping scenario~\cite{Frenklakh:2023pwy}. In contrast, the valence quark model predicts a significantly larger rapidity slope ($\alpha \approx 2.5$) in the photon-induced process, as demonstrated in Lund-based Monte Carlo studies referenced in Ref.~\cite{Lewis:2022arg}. The future Electron-Ion Collider (EIC) offers the ideal environment to test these predictions, as it will provide access to these processes (\(\gamma^*+p/A\)) via \(e+A\) and \(e+p\) collisions with unparalleled kinematic control~\cite{Frenklakh:2023pwy, Magdy:2024dpm}. Prior to the EIC’s operation, similar measurements can be performed by triggering ultra-peripheral heavy ion collisions (UPCs) at RHIC, as recently explored by the STAR collaboration~\cite{STAR:2024lvy}.

Turning to high-energy nucleus-nucleus collisions, a puzzle has long persisted. In these collisions, a large baryon excess is observed at mid-rapidity. This phenomenon contradicts the expected transport behavior of valence quarks~\cite{Gyulassy:1997mz, NA49:1998gaz, BRAHMS:2009wlg, STAR:2008med, STAR:2017sal, ALICE:2010hjm, Lacey:2024bcm}. The expected behavior is based on the following argument. In nucleus-nucleus collisions, baryon transport is expected to occur by stopping a colliding baryon from the projectile nucleus via interaction with the target nucleus’s field (see Fig.~\ref{fig:00x}). In the case of valence quarks, which carry a significant fraction (\(x_v\)) of the baryon momentum (\(P\)), their shorter interaction times (\(\Delta t_v \approx 1/(x_vP)\)) make them harder to stop by the target field. Consequently, baryon transport via valence quark stopping tends to occur closer to the fragmentation regions (i.e., at \(y \gg 0\)), far from mid-rapidity (\(y \approx 0\))~\cite{Kharzeev:1996sq}. However, this is not the case in the baryon-junction scenario. Since junctions consist of low-momentum gluons (\(x_g \ll x_v\)), they can interact for a longer duration (\(\Delta t_g \gg \Delta t_v\)) with the soft parton field of the projectile, leading to their stoppage at mid-rapidity. This argument naturally explains the large excess of baryon transport observed near \(y \approx 0\) and suggests an important consequence for electric charge transport. In the case of valence quark stopping, the baryon (\(B\)) and electric charge (\(Q\)) are strongly correlated at any rapidity, so that on average, one expects \(\langle B \rangle/\langle Q \rangle = A/Z\), 
where \(A\) and \(Z\) are the mass and atomic numbers of the colliding nuclei. In contrast, in the baryon junction scenario, the stopped junction at mid-rapidity typically forms a baryon by randomly acquiring new quarks (\(u,\, s,\, d\)) from the vacuum. As a result, the original correlation between baryon number and electric charge is lost, leading to deviations from the expected \(\langle B \rangle/\langle Q \rangle = A/Z\) behavior. 
Consequently, new studies have raised the question: ``Can the correlation between stopped baryons and charges be used to distinguish between the two baryon transport carriers scenarios?"~\cite{Lewis:2022arg}. 

However, measuring the correlation between baryon and charge transport directly is challenging because electric charge transport is much smaller than baryon transport, and various experimental uncertainties complicate a direct determination of \(\langle B \rangle/\langle Q \rangle\). 
Therefore, to improve sensitivity, it has been suggested~\cite{Lewis:2022arg} that examining the relationship between stopped baryons and charges in collisions of isobars (same $A$, different $Z$) could serve as a tool to test this correlation:
\begin{equation}\label{Eq:2}
    R({\rm Isobar}) = (B \times \Delta Z/A)/ \Delta Q,
\end{equation}
where $B$ gives the net-baryon, $Q$ is the net-charge, and $\Delta$ denotes the difference between quantities in the two isobars.
Consequently, the value of \(R({\rm Isobar})\) equal to unity (or less than unity) at mid-rapidity can be interpreted as a signature of the valence quark-stopping scenario. In contrast, a value of \(R({\rm Isobar})\) greater than unity at mid-rapidity may indicate the stronger baryon transport than electric charge, consistent with the junction-stopping scenario~\cite{Lewis:2022arg, Lv:2023fxk}. 

Motivated by such tests and expectations, a recent publication by the STAR Collaboration~\cite{STAR:2024lvy} reported measurements from isobar collisions (Ru+Ru and Zr+Zr at $\sqrt{s_{NN}}=200$ GeV) and photon-induced ($\gamma+Au$) processes that provide fresh insight into the longstanding question of whether baryon number is traced by valence quarks or a gluon junction. Together with the observed net-proton density versus beam rapidity in earlier Au+Au data, these results cannot be simultaneously described by models relying solely on conventional valence-quark expectations; instead, they appear to be more naturally accommodated within the baryon-junction picture.

Given these compelling experimental results, it is crucial to revisit the baseline within the valence-quark framework using the most advanced modeling of hadronization, fragmentation, and beam remnants. We therefore employ the state-of-the-art Lund-based PYTHIA-8 event generator (including Angantyr for heavy-ion collisions), which includes sophisticated treatments of color reconnection and junction-like structures during baryon formation; key developments that impact baryon transport\cite{Lonnblad:2023stc, Bierlich:2018xfw}. In this work, we revisit the two proposed tests in this improved framework, producing predictions that can ultimately be compared with recent STAR measurements\cite{STAR:2024lvy}.  
The article is organized as follows. Section~\ref{sec:2} introduces the PYTHIA-8 model and the data sets created. The results for the net-baryon distributions (Eq.~\eqref{Eq:1}) and the isobaric ratios (Eq.~\eqref{Eq:2}) will be presented in Section~\ref{sec:3}. Finally, a summary is provided in Section~\ref{sec:4}.

\section{Simulation Model -- PYTHIA-8 and Angantyr}\label{sec:2}
\label{sec:PYTHIA}
PYTHIA~\cite{Sjostrand:2014zea, Bierlich:2018xfw} is a widely used event generator for studying proton-proton and proton-lepton collisions at high energy. In \( p + p \) collisions, multi-parton interactions (MPI) are generated, assuming that each partonic interaction is mainly independent. PYTHIA-8 initially did not support heavy-ion systems. The Angantyr model~\cite{Bierlich:2018xfw} extends PYTHIA-8's capabilities to heavy-ion collisions, allowing for the study of $\gamma$-proton ($\gamma+p$), proton-nucleus (\(p+A\)), and nucleus-nucleus (\(A+A\)) collisions. Angantyr combines multiple nucleon-nucleon collisions into a single heavy-ion collision. It incorporates various theoretical models for producing hard and soft interactions, initial and final-state parton showers, particle fragmentations, multi-partonic interactions, color reconnection mechanisms, and decay topologies. However, Angantyr does not include any mechanism for modeling the QGP medium believed to be formed in \(A+A\) collisions.

In the current version of the PYTHIA-8 Angantyr model~\cite{Bierlich:2018xfw}, a heavy-ion collision is simulated as a series of nucleon-nucleon interactions, with each projectile nucleon potentially interacting with multiple target nucleons. The number of participating nucleons is determined using the Glauber and Glissando models~\cite{Glauber:1955qq, Rybczynski:2013yba}. Angantyr includes several algorithms to differentiate between various types of nucleon-nucleon interactions, such as elastic, diffractive, and absorptive. This model is designed to accurately describe final-state properties, including multiplicity and transverse momentum distributions, in \( A +A \) collisions.

Of relevant interest in this work is the PYTHIA-8 ability to describe hadronization and baryon number transport in high-energy collisions. In PYTHIA-8, those aspects are described using sophisticated models of color flow and color reconnection.
\emph{Color Flow} describes how QCD color charges are arranged and propagated among quarks and gluons, both before and after a collision. In PYTHIA~8, the CF determines how the proton’s three valence quarks (uud) are connected via color strings or clusters, which ultimately fragment to form hadrons.
\emph{Color reconnection} is the process, implemented after the parton shower but before hadronization, by which the color connections between partons are rearranged to reflect non-perturbative QCD effects. In PYTHIA~8, CR plays a central role in baryon formation and transport. For example, CR may relocate the color junction associated with the proton’s baryon number, or even produce new junction--antijunction pairs, leading to the formation of baryons and antibaryons away from the beam remnants.

In this work, we aim to systematically assess the effects of different CF and CR models on the baryon dynamics in heavy-ion collisions by understanding their impact on observables such as the isobar R(Isobar) ratio and the net-proton rapidity slopes in isobar and photonuclear processes. Consequently, we will present our calculation for two CF and three CR schemes described as:

\subsubsection{CF-1 using Legacy Leading-Order Parton Shower (LO PS) and CR-1 using MPI-based model}
In the default scenario of older versions of PYTHIA, the CF is constructed using the \emph{leading color approximation} ($N_c \to \infty$). Within this framework, each parton shower and each Multiple Parton Interaction (MPI) system is evolved and color-connected independently, such that final-state partons are arranged in simple dipole chains. The so-called ``MPI-based'' color reconnection model is then applied, attempting to reconnect color lines between separate MPI systems, typically by swapping color tags in a way that minimizes the overall string length. This method provides a mechanism for generating color correlations between partons originating from different sub-collisions~\cite{Sjostrand:2004pf, Christiansen:2015yqa}.

In this scenario, baryon formation is primarily achieved through diquark-antidiquark pair production during the Lund string fragmentation process. The Y-shaped Junctions topologies that can efficiently facilitate baryon creation are rare, since their formation is suppressed in the leading color limit and not explicitly enhanced by the MPI-based CR model. Consequently, hadronization proceeds by fragmenting the dipole-connected string pieces into hadrons, with baryon production being suppressed and closely linked to the available phase space for diquark formation. As a result, baryon number transport is largely limited; the baryon number typically follows the beam remnants, unless reconnections happen to bring a string endpoint (such as a valence quark or diquark) into the central rapidity region. This scenario, therefore, struggles to account for the significant baryon enhancement and the broad baryon number transport that have been observed in high-multiplicity events at the LHC~\cite{Christiansen:2015yqa}.

\subsubsection{CF-1 using LO PS amd CR-2 using Gluon-move model}
In this scenario, the leading color (LC) parton shower is still used to establish the initial color connections between partons. The gluon-move'' CR model, implemented in PYTHIA-8 as a toy model, extends the possibilities of CR by allowing the color-anticolor indices of gluons to be swapped or moved'' along dipole chains. While the primary objective remains to minimize the total string length, this approach permits more extreme modifications to the color topology than the MPI-based model, as entire gluons can change their position in the color chain. Nonetheless, the model remains confined to the dipole picture and does not fully incorporate subleading color effects, nor does it allow for the explicit formation of baryonic (junction) structures.

Baryon formation in this scenario continues to occur mainly through diquark-antidiquark creation during Lund string fragmentation, as in the MPI-based model. The increased flexibility in rearranging gluons can indirectly enhance baryon production by lowering the invariant mass of string pieces, thereby expanding the available phase space for diquark formation. Hadronization is still governed by the Lund string model, but the additional rearrangement freedom introduced by the gluon-move model can result in slightly shorter and more boosted string systems, which in turn affect the properties of the produced hadrons. However, baryon number transport is only marginally affected; while this model enables greater mixing among string pieces, it does not efficiently generate junctions or facilitate significant baryon number transport into the central rapidity region~\cite{Christiansen:2015yqa}.

\subsubsection{CF-2 using Full SU(3) Color Flow Model and CR-3 using Advanced SU(3) Model}
This scenario represents the most advanced and theoretically complete approach currently available in PYTHIA-8~\cite{Christiansen:2015yqa, Lonnblad:2023stc}. Within this framework, color flow is modeled using an approximate \emph{full SU(3) color algebra} treatment, extending beyond the traditional leading color approximation. Each parton—whether a quark or a gluon—is assigned a color index, and the model systematically includes subleading color configurations in accordance with the true SU(3) group structure. The ``Advanced SU(3)'' color reconnection model utilizes these richer color assignments, considering not only simple dipole reconnections but also the dynamic formation of \emph{junctions}—Y-shaped topologies associated with baryonic color structures, as well as more general multi-parton color connections. The algorithm iteratively searches for configurations that minimize the total string potential energy or string length, thereby facilitating the natural emergence of baryonic structures and enhancing color coherence throughout the event.

Baryon formation in this scenario is significantly increased through two key mechanisms: the standard production of diquarks during string breaking, and the explicit \emph{junction formation} enabled by advanced color reconnection. Junctions, in particular, provide a natural origin for baryons by allowing three quarks—potentially originating from different MPI systems—to combine at a single baryon vertex. Hadronization proceeds via an extended version of the Lund string model that can accommodate these complex junction topologies, with fragmentation occurring along the multiple legs of the Y-shaped configuration. This leads to nontrivial kinematic distributions and enables the production of baryons with high transverse momentum, closely matching experimental observations.

Most notably, baryon number transport is greatly enhanced in this model. The baryon number, carried by the string junction, can be transported over large rapidity intervals and is no longer restricted to the beam remnants. As a result, the model successfully reproduces the observed baryon-to-meson ratios and the scaling of baryon production with event multiplicity, thereby providing a unified and consistent description across $e^+e^-$, pp, and heavy-ion collision data~\cite{Christiansen:2015yqa, Lonnblad:2023stc}.

\section{RESULTS}\label{sec:3}
In this section, we will discuss the observables: (i) the rapidity slope of the net-baryon distributions for $\gamma_{\rm e}$+ \( p \) collisions and (ii) the isobaric ratio given by Eq.~\eqref{Eq:2} for Ru+Ru and Zr+Zr collisions at 200 GeV.

Photon-induced interactions on proton and nuclei $\gamma$ + \( (p)A \) have been proposed as an effective setting for studying baryon stopping for two key reasons.  First, the photon is the simplest probe and, to first order, can fluctuate into a single dipole, interacting with the carriers of baryon number, providing a controlled way to study its stopping dynamics.
For example, at RHIC \( Au+Au \) UPC ($\sqrt{s_{_{NN}}}=54.4$ GeV), photons with energies up to 2 GeV~\cite{STAR:2024lvy} are used to test whether the baryon carrier inside 27.2 GeV nucleons can be stopped at midrapidity. These low-energy photons may lack the stopping power to stop high-momentum valence quarks but may effectively halt low-momentum gluons that form the baryon junction, thereby producing a baryon at midrapidity. 
Second, the experimental advantage is that detectors like STAR at RHIC have excellent baryon identification capabilities only in the midrapidity region. Since a projectile photon contains no baryons, \(\gamma+(p)A\) interactions yield a clearer observation of the characteristic functional shape of baryon stopping from the target, facilitating direct comparison with predictions from either valence quark or baryon junction scenarios. In contrast, hadronic \(p+p\) or \(A+A\) collisions involve overlapping distributions from baryons in both the target and projectile, resulting in a flatter rapidity slope that is more challenging to measure with limited midrapidity acceptance~\cite{Lewis:2022arg}. 
\begin{figure}[!h]
\includegraphics[width=1.0  \linewidth, angle=-0,keepaspectratio=true,clip=true]{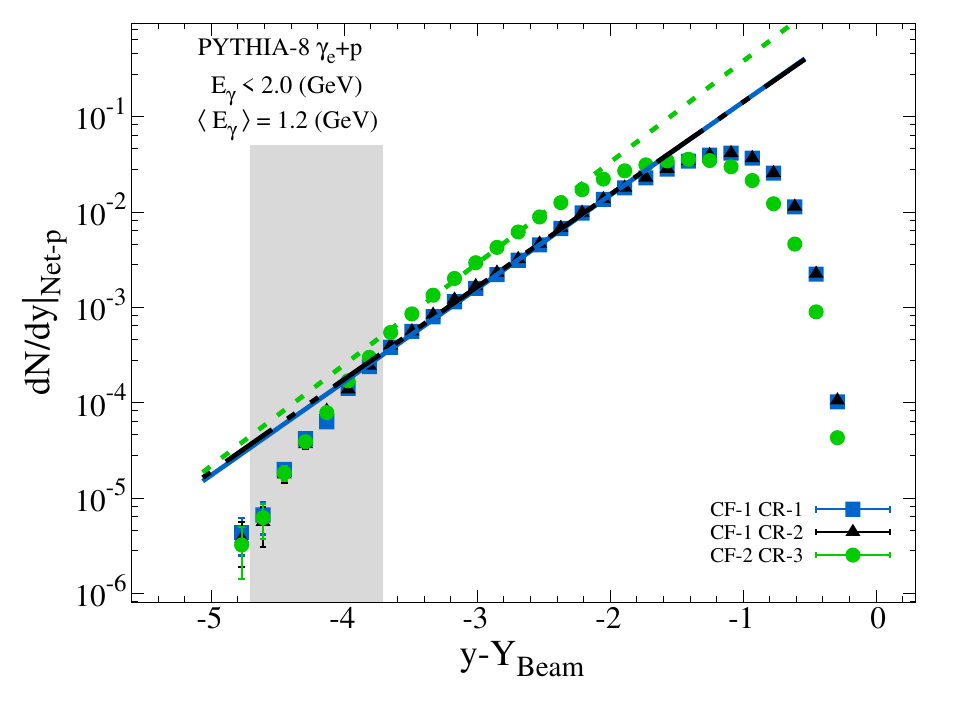}
\vskip -0.4cm
\caption{
The net-baryon rapidity distributions for $\gamma_{\rm e}$+p collisions, corresponding to the different PYTHIA-8 modes, are shown. The lines represent exponential fits to the PYTHIA-8 calculations in the rapidity range $y - Y_{\text{Beam}} = -2.5$ to $-5.0$. 
\label{fig:fig1}}
\vskip -0.3cm
\end{figure}

The predicted rapidity slope of the net-baryon distribution, $\alpha_{B}$, is the primary distinguishing feature between the transport implementations in quark-based and baryon junction models. We, therefore, simulate \(\gamma_{\rm e}+p\) collisions where \(\gamma_{\rm e}\) denotes a photon emitted by an electron with an energy of 5~GeV and restricted the photon energy to $E_{\gamma} < 2.0$~GeV. Figure~\ref{fig:fig1} shows the net-baryon rapidity distributions for $\gamma_{\rm e}$+p collisions as simulated by the PYTHIA-8 model with various implementations of baryon stopping mechanisms. Away from the beam-fragmentation region, the net-baryon yield exhibits an approximate exponential dependence on rapidity; however, the entire kinematic range cannot be described by a single exponential function. The slope from exponential fits, extracted over the rapidity range $y - Y_{\text{Beam}} = -2.5$ to $-5.0$, depend sensitively on the PYTHIA-8 mode: we obtain $\alpha_{B} = 2.253 \pm 0.05$ for CF-1 CR-1,  $\alpha_{B} = 2.231 \pm 0.06$ for CF-1 CR-2, and $\alpha_{B} = 2.443 \pm 0.04$ for CF-2 CR-3. The observed variation indicates a moderate sensitivity of baryon stopping to the treatment of CF and CR in PYTHIA-8, which is in agreement with previous calculations~\cite{Lewis:2022arg}. The weak sensitivity to the CF and CR schemes in $\gamma_{\text{re}}+p$ collisions reflects the absence of a dense color environment and the lack of multiple parton interactions. Even with advanced color reconnection mechanisms, such as CF-2, there is no significant impact on the rapidity transport of baryon number~\cite{Kharzeev:1996sq, Vance:1998vh, Skands:2010ak, Lonnblad:2023stc}. Consequently, in $\gamma_{\text{re}}+p$ collisions as modeled in PYTHIA-8, the baryon number remains predominantly associated with the proton remnant.

\begin{figure}[!h] 
\includegraphics[width=1.0  \linewidth, angle=-0,keepaspectratio=true,clip=true]{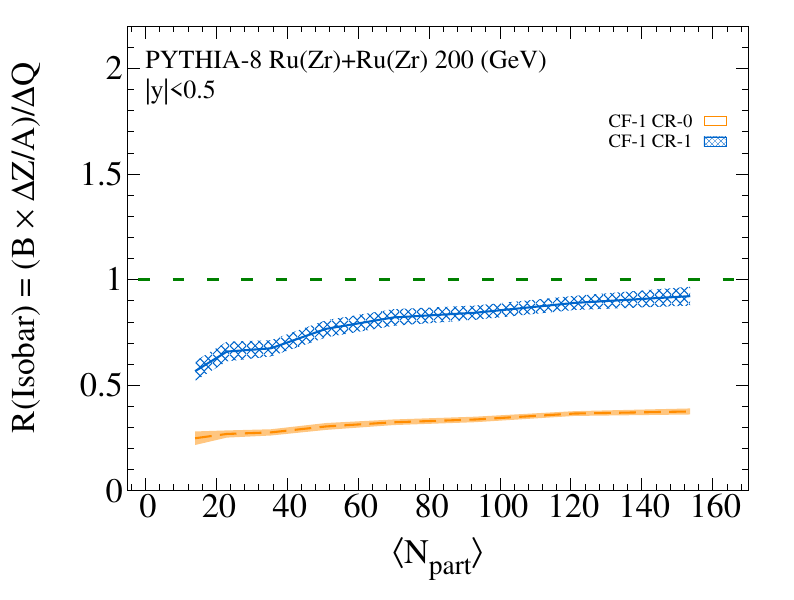}
\vskip -0.4cm
\caption{
The R(Isobar) as a function of $N_{\rm part}$ at 200 GeV from the PYTHIA-8 model with (i) $CF-1$ $CR$-Off ($CR-0$) and (ii) $CF-1~CR-1$ schemes. 
}\label{fig:fig3}
\vskip -0.0cm
\end{figure}
\begin{figure}[!h] 
\includegraphics[width=0.99  \linewidth, angle=-0,keepaspectratio=true,clip=true]{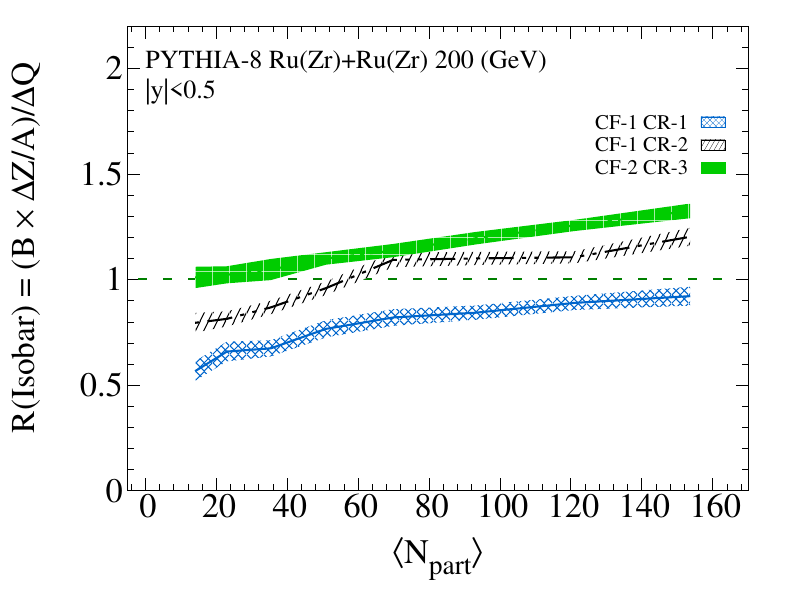}
\vskip -0.4cm
\caption{
The same as Fig.~\ref{fig:fig3} but for different PYTHIA-8 schemes given in Sec.~\ref{sec:PYTHIA}. 
}\label{fig:fig4}
\vskip -0.0cm
\end{figure}

We now move on to explore the relationship between charge and baryon number multiplicity, which presents a potential method to distinguish whether valence quarks or baryon junctions are the primary carriers of baryon numbers. However, accurately measuring the net charge yield is challenging because it involves determining the small net charge from the minimal difference between the large yields of positively and negatively charged particles~\cite{Lv:2023fxk}. To overcome this difficulty, it has been proposed to evaluate the net charge difference between isobaric collisions of $_{40}^{96}\mathrm{Zr}$ + $_{40}^{96}\mathrm{Zr}$ and $_{44}^{96}\mathrm{Ru}$ + $_{44}^{96}\mathrm{Ru}$~\cite{Lewis:2022arg}. In these collisions, the net-charge difference ($\Delta Q = Q_{\rm{Ru}}-Q_{\rm{Zr}}$) can be calculated using double ratios by comparing the positive and negative particles and then contrasting Ru+Ru collisions with Zr+Zr collisions. The key advantage of this isobaric method is that the use of double ratios cancels many systematic effects, thereby enabling a more precise measurement of the net charge yield. In addition, the measurement of the net-baryon yield, $B$, which should be identical in the two systems, is relatively straightforward. This allows for the computation of the ratio $R({\rm Isobar})$ as shown in Eq.~\eqref{Eq:2}. In the valence-quark stopping picture, $\Delta Q$ should approximate $B \times \Delta Z/A$, resulting in $R({\rm Isobar}) \approx 1.0$ at mid-rapidity. Conversely, if baryon-junction stopping is prevalent, $\Delta Q$ will be less than $B \times \Delta Z/A$, leading to $R({\rm Isobar}) > 1.0$ at mid-rapidity. In this study, we provide baseline calculations of $R({\rm Isobar})$ for $_{40}^{96}\mathrm{Zr}$ + $_{40}^{96}\mathrm{Zr}$ and $_{44}^{96}\mathrm{Ru}$ + $_{44}^{96}\mathrm{Ru}$ using the PYTHIA-8 model framework, with the several CF and CR scenarios. For these simulations, PYTHIA-8 is run in its Angantyr mode, which extends its applicability to heavy-ion collisions~\cite{Bierlich:2018xfw}.

Figure~\ref{fig:fig3} shows $R({\rm Isobar})$ as a function of the number of participating nucleons ($N_{\rm part}$) for Ru(Zr)+Ru(Zr) collisions at 200 GeV, comparing the PYTHIA-8 model results with the CF-1 scheme and with the CR scheme turned on and off. Both simulations, performed at mid-rapidity, exhibit $R({\rm Isobar})$ values below unity that decrease as $N_{\rm part}$ decreases. Furthermore, our simulations indicate that the CR-off (CR-0) configuration produces significantly lower $R({\rm Isobar})$ values, while enabling the CR (CR-1 scheme) increases $R({\rm Isobar})$ toward unity, although it remains below one across all centralities. In contrast, STAR measurements report $R({\rm Isobar}) > 1.0$ throughout the entire $N_{\rm part}$ range~\cite{STAR:2024lvy}, which has been suggested as a distinguishing signature of baryon junction dynamics. The behavior observed in PYTHIA-8 underscores the experimental results, reflecting the absence of non-planar color topologies in the CF-1 CR-1 scheme. This limitation causes the baryon number to remain localized to the beam remnants, resulting in limited baryon transport to the mid-rapidity region.

The PYTHIA-8 model offers several CF and CR schemes, as summarized in Sec.\ref{sec:PYTHIA}. The CF-1 CR-1 and CF-1 CR-2 schemes both lack non-planar color topologies (e.g., color junctions), resulting in the localization of baryon number to the beam remnants. In contrast, the CF-2 CR-3 scheme permits the simulation of non-planar color topologies, enabling the transport of baryon number to mid-rapidity. These features are illustrated in Fig.\ref{fig:fig4}, which compares $R({\rm Isobar})$ as a function of $N_{\rm part}$ for the CF-1 CR-1, CF-1 CR-2, and CF-2 CR-3 schemes. As discussed earlier, the CF-1 CR-1 scheme yields values below unity, while both the CF-1 CR-2 and CF-2 CR-3 schemes produce results that are 25–30\% above unity in central collisions. Our findings indicate that even in the absence of color junctions, a sophisticated CR mechanism, as implemented in CR-2, can enhance baryon production at mid-rapidity in central collisions, resulting in $R({\rm Isobar}) > 1$. In contrast, the CF-2 CR-3 scheme, which incorporates color junctions, yields $R({\rm Isobar}) > 1$ across all centralities presented. These results suggest that advanced CF and CR schemes in PYTHIA-8, particularly those including color junction features, shift the model predictions closer to experimental observations~\cite{STAR:2024lvy}. This outcome aligns with the qualitative expectations from baryon junction dynamics and highlights the need for future quantitative comparisons with STAR data.

\begin{figure}[htb] 
\includegraphics[width=1.0  \linewidth, angle=-0,keepaspectratio=true,clip=true]{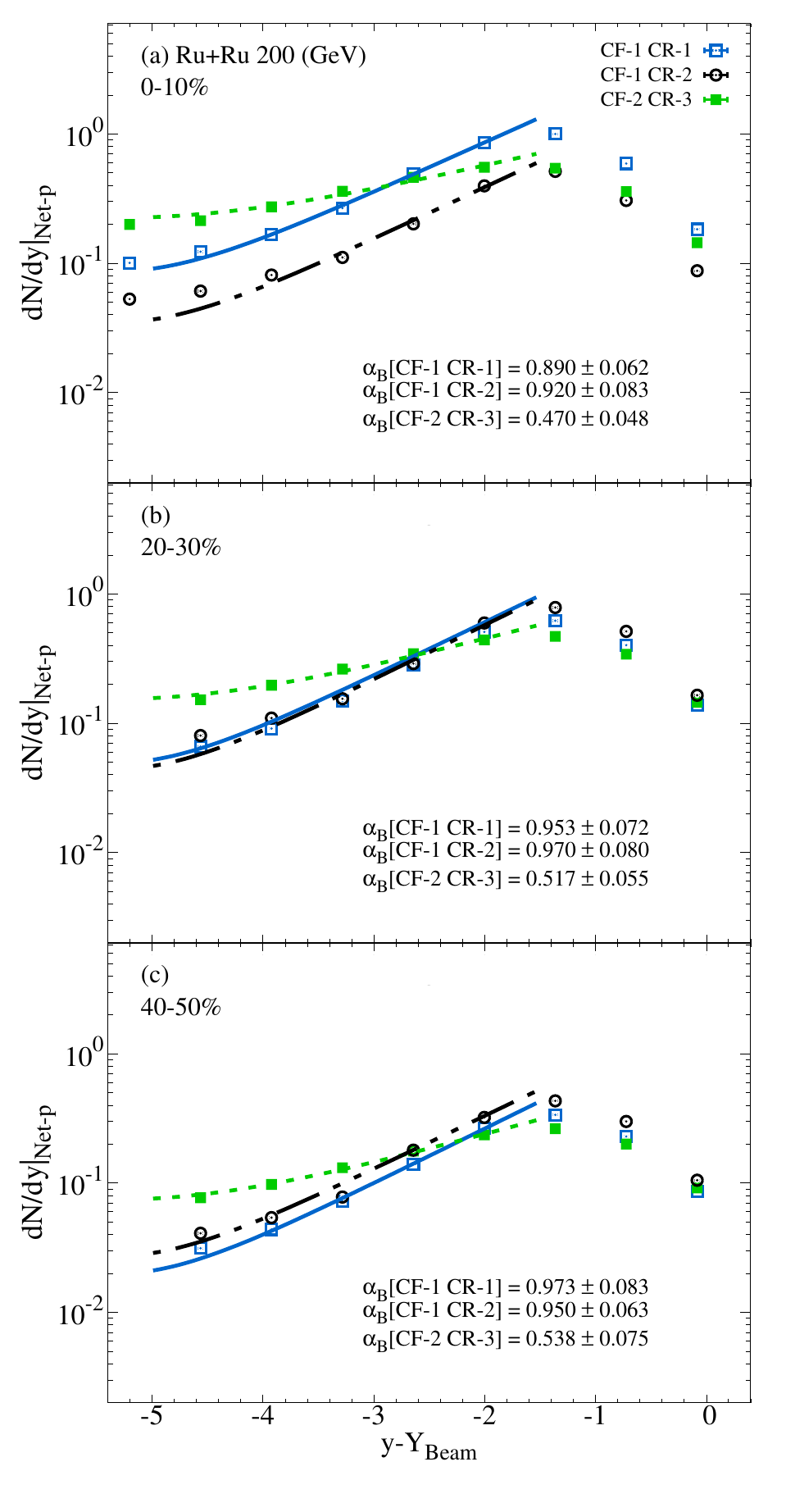}
\vskip -0.4cm
\caption{
The scaled net-baryon rapidity distributions for Ru+Ru collisions at 200~GeV, obtained from the PYTHIA-8 model under the configuration in Sec.~\ref{sec:PYTHIA}, are shown for three centrality selections. The lines represent exponential fits to the PYTHIA-8 calculations. Because the system is symmetric, only the positive rapidity range is displayed.
}\label{fig:fig5}
\vskip -0.3cm
\end{figure}

The results shown in Fig.~\ref{fig:fig4} demonstrate that both the CF-1 CR-2 and CF-2 CR-3 schemes, which utilize different CF models, yield $R({\rm Isobar})$ values greater than unity. This finding prompts the question of whether it is possible to further distinguish between the CF models beyond just the $R({\rm Isobar})$ calculations. A key distinction between the CF-1 and CF-2 models lies in the locality of the baryon number: the presence of color junctions in CF-2 enables the transport of baryon number across large rapidity intervals. To explore this feature, we propose examining the rapidity dependence of net-proton density in Ru+Ru collisions at 200~GeV, as illustrated in Fig.\ref{fig:fig5}. Given that we expect a similar rapidity distribution for baryons in Zr+Zr collisions, a separate study for that system was not conducted. Symmetric A+A collisions typically produce a nearly symmetric rapidity distribution of net baryons, as both the target and projectile contribute equally. The baryon stopping from each nucleus results in an exponential decrease in net-baryon yields near mid-rapidity ($y = 0$). Following the expectation outlined in Ref.~\cite{Kharzeev:1996sq}, this behavior can be parameterized in terms of rapidity loss variable $y-Y_{\text{Beam}}$: 
\begin{equation}\label{Eq:AA}
\frac{dN_{p-\bar{p}}}{dy} = N_B \left( e^{-\alpha_B (y - Y_{\text{Beam}})} + e^{-\alpha_B (-y - Y_{\text{Beam}})} \right),
\end{equation}
where $N_B$ and $\alpha_B$ are the parameters that can be obtained from a fit.

The calculations presented in Fig.\ref{fig:fig5} illustrate the exponential dependence of net-proton density at mid-rapidity for Ru+Ru collisions at 200~GeV. These results include all the versions of the PYTHIA-8 model in Angantyr mode for configurations listed in Sec.~\ref{sec:PYTHIA}. Three centrality classes (0–10\%, 20–30\%, and 40–50\%) are displayed, with only the positive rapidity region shown due to the system's symmetry. The lines in each panel represent exponential fits of the form given in Eq.~\eqref{Eq:AA}, from which the slope parameter \(\alpha_B\) is extracted. This exponential decrease reveals a consistent trend with varying slopes across different PYTHIA-8 CF and CR schemes, indicating a common underlying baryon-stopping mechanism. Our simulations demonstrate that the indicated exponential pattern remains independent of the CR setting for the same CF model. As observed, CF-1 CR-1 and CF-1 CR-2 schemes, both employing the legacy CF model, yield similar \(\alpha_B\) values across different centrality selections, while the CF-2 CR-3 scheme, based on the full SU(3) CF model, produces a slope that is approximately 0.5 lower. This distinct variation underscores the sensitivity of the net‐baryon rapidity distribution to the underlying CF mechanisms, highlighting its potential as a stringent test for constraining PYTHIA‑8 model parameters and differentiating between distinct CF schemes. Although isobar rapidity slope measurements have not yet been presented~\cite{STAR:2024lvy}, future data-model comparisons promise to provide important insights.

\section*{Conclusions and Outlook}~\label{sec:4}
In this work, we have investigated two key aspects of baryon transport in high-energy nuclear collisions using the PYTHIA-8 event generator with multiple CF and CR schemes: (i) the rapidity slope of net-baryon distributions, and (ii) the baryon-to-net-charge double ratio $R({\rm Isobar})$ in isobaric Ru+Ru and Zr+Zr collisions. Our study was motivated by recent STAR measurements~\cite{STAR:2024lvy} that address the longstanding question of what mechanism tracks baryon number transport in high-energy collisions~\cite{Kharzeev:1996sq}.

Our calculations for $\gamma_{\rm e}+p$ collisions show that the net-baryon distributions exhibit an approximately exponential dependence on rapidity, with extracted slopes depending weakly on the particular CF and CR mechanisms used in PYTHIA-8. This weak sensitivity reflects the unique features of photon-induced processes, such as the absence of a dense color environment and multiple parton interactions. Even the most advanced color reconnection models in PYTHIA-8 do not significantly alter baryon number transport in these systems; the baryon number remains mostly associated with the proton remnant, with minimal transport into central rapidity. 

For isobaric heavy-ion collisions, our analysis reveals important distinctions between CF and CR schemes. The default PYTHIA-8 configuration (CF-1 CR-1), which lacks junction structures, yields $R({\rm Isobar})$ values below unity, indicating limited baryon transport to mid-rapidity. The CF-1 CR-2 scenario, with the standard CF but a more advanced CR model, enhances baryon production at mid-rapidity, resulting in $R({\rm Isobar}) > 1$ in central collisions. Similarly, the CF-2 CR-3 scheme—featuring an advanced CF model with explicit color junctions and sophisticated CR—enables baryon number transport across rapidity and also produces $R({\rm Isobar}) > 1$. Importantly, we find that $R({\rm Isobar})$ alone does not distinguish between CF-1 CR-2 and CF-2 CR-3; both configurations achieve similar qualitative enhancements. This motivates our proposal to use the net-baryon rapidity distribution in Ru+Ru collisions as a complementary observable, which is more directly sensitive to the presence of color junctions and the mechanism of baryon number transport across rapidity, providing a path to further discriminate between these underlying CF models.

In summary, motivated by ongoing experimental efforts to clarify the role of valence quarks and junction structures in baryon number transport, our study systematically explores how different CF and CR mechanisms in PYTHIA-8 affect key observables related to baryon stopping. While recent model developments allow a more realistic treatment of baryon transport, especially at mid-rapidity, our findings highlight both the progress made and the challenges that remain. Continued theoretical advances, together with high-precision experimental data, will be crucial for a comprehensive understanding of baryon transport in high-energy nuclear collisions. We anticipate that further quantitative comparisons and new measurements, particularly those sensitive to rapidity-dependent baryon transport, will play a decisive role in unraveling the mechanisms underlying baryon stopping in QCD.

\section*{Acknowledgments}
We thank David Frenklakh, Rongrong Ma, Dmitri Kharzeev, Nicole Lewis, Leif L\"onnblad, Zebo Tang, and Zhoudunming Tu for the valuable discussions and acknowledge the attendance and discussion at the CFNS workshop titled "The 1st workshop on Baryon Dynamics from RHIC to EIC". This work was supported in part by the Offices of NP within the U.S.\ DOE Office of Science under the contracts DE-FG02-89ER40531 and DE-SC0012704.
This work was performed by using computational resources offered by the High-Performance Computing Center at Texas Southern University.

\bibliography{ref} 
\end{document}